\pdfoutput=1

\documentclass[11pt]{article}

\usepackage[final]{acl}

\usepackage{times}
\usepackage{latexsym}

\usepackage{amsmath}
\usepackage{amssymb}
\usepackage{mathtools}
\usepackage{amsthm}
\usepackage{adjustbox}

\usepackage{subfigure}
\usepackage{booktabs}
\usepackage{multirow}
\usepackage{multicol}
\usepackage{makecell}
\usepackage{color, colortbl}
\usepackage[capitalize,noabbrev]{cleveref}

\usepackage[T1]{fontenc}

\usepackage[utf8]{inputenc}

\usepackage{microtype}

\usepackage{inconsolata}

\usepackage{graphicx}

\title{MultiVerse: Efficient and Expressive Zero-Shot Multi-Task Text-to-Speech}

\author{
 \textbf{Taejun Bak\textsuperscript{1}\thanks{Work performed at NCSOFT.}},
 \textbf{Youngsik Eom\textsuperscript{2}},
 \textbf{Seungjae Choi\textsuperscript{2}},
 \textbf{Young-Sun Joo\textsuperscript{2}},
\\
 \textsuperscript{1}SK Telecom, Republic of Korea \\
 \textsuperscript{2}Audio AI Lab., NC Research, NCSOFT Corp., Republic of Korea
\\
  \texttt{taejun.bak@sk.com} \\
  \texttt{\{yseom, seung, ysjoo555\}@ncsoft.com} \\
}

\begin{document}
\maketitle
\begin{abstract}
Text-to-speech (TTS) systems that scale up the amount of training data have achieved significant improvements in zero-shot speech synthesis. However, these systems have certain limitations: they require a large amount of training data, which increases costs, and often overlook prosody similarity. To address these issues, we propose MultiVerse, a zero-shot multi-task TTS system capable of performing TTS and speech style transfer in zero-shot and cross-lingual conditions, while requiring much less training data than traditional data-driven approaches. To ensure zero-shot performance even with limited data, we leverage source-filter theory-based disentanglement, utilizing the prompt for modeling filter-related and source-related representations. Additionally, to further enhance prosody similarity, we adopt a prosody modeling approach combining prompt-based autoregressive and non-autoregressive methods. Evaluations demonstrate the remarkable zero-shot multi-task TTS performance of MultiVerse and show that MultiVerse not only achieves zero-shot TTS performance comparable to data-driven TTS systems with much less data, but also significantly outperforms other zero-shot TTS systems trained with the same small amount of data. In particular, our innovative prosody modeling technique enables Multiverse to generate speech with a high degree of prosody similarity to the given prompts. Our samples are available at \url{https://nc-ai.github.io/speech/publications/multiverse/index.html}
\end{abstract}

\section{Introduction}
\label{Introduction}
Deep learning-based text-to-speech (TTS) has advanced to the point of synthesizing human-like speech \cite{wang2017tacotron, ren2019fastspeech}. However, recent research has been extended beyond this scope, exploring various ways of broadening the application of speech synthesis models. Representative tasks include synthesizing unseen speaker’s speech, known as \textbf{zero-shot TTS} \cite{jia2018transfer, cooper2020zero}, generating speech in a language that the monolingual target speaker has not seen, referred to as \textbf{cross-lingual TTS} \cite{cho22_interspeech, zhang19e_interspeech, xin2021disentangled}, and transferring the prosody of a speech reference to a target speaker, known as \textbf{speech style transfer} \cite{lee21h_interspeech, huang2022generspeech}. Furthermore, recent TTS research has integrated the zero-shot task with cross-lingual or style transfer tasks \cite{casanova2022yourtts, zaidi22b_interspeech}.

To expand TTS applications in zero-shot conditions, it is crucial to ensure generalization across various speech components, such as content, style, and speaker identity. Disentangled representations facilitate this by enabling the system to capture interpretable and controllable features, thereby improving generalization of TTS systems through the learning of these individual components \cite{rpr-learning, mythos}. However, due to the often entangled nature of these components, separately learning their general characteristics remains challenging.

\begin{figure*}[t]
  \centering
  \centerline{\includegraphics[width=1.0\linewidth]{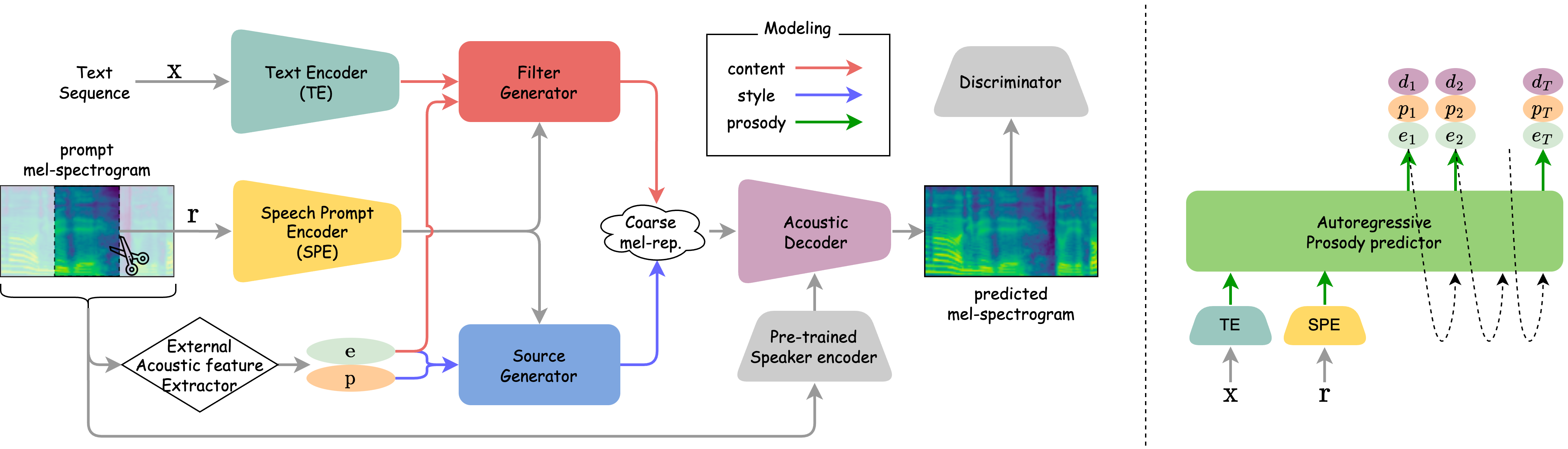}}
  \vskip -0.1in
  \caption{Overall structure of MultiVerse. The acoustic model and the autoregressive prosody predictor are on the left and right side of the figure, respectively. During training, overall modules are trained together, except the pre-trained speaker encoder. Multi-task TTS can be accomplished by varying input conditions.}
  \label{fig:ditto}
\end{figure*}

Data-driven methods are a representative approach to learning generalized acoustic components from large-scale speech datasets. Specifically, these methods scale up the TTS system to train on massive datasets, making them robust under unseen conditions \cite{wang2023valle, zhang2023vallex, voicebox}. Additionally, \citet{jiang2023mega, jiang2024megatts, shen2023naturalspeech, ju2024naturalspeech} adopt disentangled modeling to separately learn acoustic components. Disentangled modeling contributes to ensure generalization in each component by independently encapsulating interpretable elements. However, a significant amount of training dataset is required as the decoder needs to learn the relationships between disassembled elements. Preparing large-scale data is especially challenging for minority languages. Moreover, even with large-scale training data, there is still potential room for improvement in prosody similarity in zero-shot scenarios.

In this paper, we introduce a multi-task TTS system, called MultiVerse, enabling speech synthesis and speech style transfer in zero-shot and cross-lingual conditions, requiring significantly less data compared to the data-driven approaches and featuring enhanced prosody modeling. MultiVerse enhances training efficiency by source-filter theory-based decomposed modeling \cite{fant1970acoustic}. Specifically, MultiVerse decomposes speech generation into filter- and source-related representation generation, with prompt speech utilized in the modeling of each representation. Notably, both representations result in features that have a similar distribution to the mel-spectrogram \cite{bak21_interspeech}. Therefore, it is suitable for the decoder to learn the interdependent relationship between the representations even with small data. Furthermore, MultiVerse introduces an effective style modeling method. While several zero-shot models have modeled either the acoustic features \cite{shen2023naturalspeech} or prosody latents to capture speech style \cite{jiang2023mega}, MultiVerse utilize both autoregressive (AR) based acoustic feature modeling and Non-AR based prosody modeling.

We evaluate MultiVerse's performance in zero-shot TTS tasks under various conditions. Regarding language, we evaluate both intra- and cross-lingual speech synthesis and measure naturalness and similarity in neutral and expressive speech style. Additionally, we compare our model with a large-scale TTS model using data-driven methods and a speech style transfer model. Evaluation results demonstrate that MultiVerse has the following advantages: (1) Zero-shot intra- and cross-lingual TTS is achievable with a small amount of training data. MultiVerse can achieve similar zero-shot synthesis in both timbre and prosody with only $\frac{1}{60}$ of the training data compared to VALL-E \cite{wang2023valle}. (2) The proposed prosody modeling is also effective in reflecting conditions in various tasks of speech synthesis. Even without requiring information about the content or phoneme duration of the prompt speech, it can reflect prosody similar to the prompt in both intra-lingual and cross-lingual settings.

\section{MultiVerse}
\label{proposed_methods}
MultiVerse models speech by disassembling it into two components: filter and source. The proposed model comprises three main modules: (1) an acoustic model based on the source-filter theory \cite{fant1970acoustic} that generates mel-spectrograms given text and speech prompts, (2) an AR prosody predictor that predicts prosody-related acoustic features (duration, pitch, energy) from input conditions, and (3) a discriminator for adversarial training.

\subsection{Source-Filter Theory Based Decomposed Modeling}
Inspired by the source-filter theory, which explains the response between the vocal tract filter and the sound source, the proposed acoustic model generates mel-spectrogram through the filter and source generators. The filter generator is tasked with producing the vocal tract filter-related representation, while the source generator outputs the source-related representation. We name these two representations as filter representation and source representation, respectively. Both representations are obtained from feed forward transformer-based generator \cite{ren2019fastspeech}.

\textbf{Filter representation}\space\space We consider filter representation as hidden states that contain information related to speech content, pronunciation, and speaker identity, but is less dependent on prosody. This representation, obtained by the filter generator, is modeled by taking phoneme representation as input, along with the energy embedding. Both input features are upsampled by the gaussian upsampling \cite{shen2020non}.

\textbf{Source representation}\space\space We consider source representation primarily as hidden states that contain prosodic information, such as intonation, rhythm, and stress patterns, with low dependence on content. The source generator produces source representation from frame-wise upsampled phoneme-level pitch and energy embeddings. During training, it generates the source representation from ground-truth acoustic feature embeddings, while during inference, it utilizes predicted acoustic feature embeddings.
We provide further experimental analysis on both representations in \cref{app:representations}.

 \citet{choi2021nansy, bak21_interspeech} also utilized source-filter based speech disentanglement. Unlike existing methods, our approach adopts prompt-based modulation in modeling both representations. Since both vocal tract filter and sound source are influenced by speaker characteristics, the prompt speech is reflected in both generators. The mel-spectrogram of prompt speech serves as the input for obtaining hidden states from the speech prompt encoder. Parameters for the FiLM \cite{film} layer are predicted from the cross-attention output between the input of generators and the output of speech prompt encoder as in \cite{shen2023naturalspeech}. Subsequently, the FiLM layer modulates the representations within the generators.

\subsection{Increasing Filter Capacity of the Acoustic Decoder}
The intermediate representation formed by combining both representations, referred to as coarse mel-representation, resembles the interaction between the vocal tract filter and the sound source \cite{bak21_interspeech}. Since this fusion follows the frequency response in the source-filter model, the coarse mel-representation is closely related to a high-dimensional features of speech. Consequently, the acoustic decoder has an advantage in learning the interdependent relationship between the filter and source representations.

To produce mel-spectrograms while preserving various types of information such as speech content and style within the coarse mel-representation, it is necessary to increase the filter capacity of the acoustic decoder. To increase the filter capacity, the acoustic decoder's transformer block replaces convolution layers with sample-adaptive kernel selection-based convolution layers \cite{gigagan}. It aims to find suitable convolution filters for the speech prompt. Specifically, learnable filters of each convolution layer are weighted sums based on predicted weights from the global style embedding. The aggregated filter is then modulated and de-modulated \cite{stylegan2}. The global style embedding is derived from the pre-trained speaker encoder. More detail information about sample-adaptive kernel selection is described in \cref{app:sample-adaptive-detail}.

\subsection{Two-stage Prosody Modeling}

MultiVerse's prosody modeling consists of two stages: first, the prosody predictor models the acoustic features autoregressively; second, the source generator models prosody in the latent space non-autoregressively using these acoustic features.

\subsubsection{Autoregressive Prosody Modeling}
\label{ar_prosody}
The proposed AR prosody predictor models the time-varying distribution of acoustic features (duration, pitch, energy) as a conditional language modeling task. The goal of the AR prosody predictor is prediction of acoustic units suitable for the given text and prompt conditions. Due to the time-dependent nature of prosody and the need to model large-variations in prosody, we adopt an AR approach \cite{kharitonov-etal-2022-text}. The prosody predictor is trained to predict acoustic units $\mathbf{c}_t=\{\mathbf{d}_t,\mathbf{p}_t,\mathbf{e}_t\}$ corresponding to phoneme sequences $\mathbf{x}=\{x_1,x_2,...,x_T\}$, where $\mathbf{d}$, $\mathbf{p}$, $\mathbf{e}$ are the duration, pitch, and energy unit sequences, respectively. These unit sequences of which each value corresponds to an index are obtained by quantizing normalized acoustic sequences. Prosody modeling, which is conditioned on the speech prompt $\mathbf{r}$ and the phoneme sequence $\mathbf{x}$, is written as the following equation:
\vskip -0.2in
\begin{equation}\label{prosody}
    p(\mathbf{c}|\textbf{x},\textbf{r};\theta_{ARP})=\prod^T_{t=0}p(\mathbf{c}_t|\mathbf{c}_{<t},\textbf{x},\textbf{r};\theta_{ARP}),
\end{equation}
where $\theta_{ARP}$ represents the parameters of AR prosody predictor. To model the prosody using a prompt-based in-context learning, we utilize the phoneme sequence and the prompt as a prefix, a similar approach to \citet{wang2023valle}. 

The AR approach is also utilized in the prosody latent language model (P-LLM) in Mega-TTS \cite{jiang2024megatts}, which autoregressively models vector-quantized codebook \cite{vq} of prosody hidden states. However, the performance of vector quantization depends on the quantity and diversity of training data \cite{gersho2012vector}. In contrast, the AR prosody predictor, which models acoustic feature units, is data-efficient. We provide further analysis on the limitation of VQ based modeling in \cref{app:pllm}. Additionally, the P-LLM is limited to specific prompt speech, such as the particular languages present in the training data, and requires alignment information of the prompt. Conversely, the AR predictor does not impose restrictions on the utilization of prompt speech.

\subsubsection{Non-Autoregressive Prosody Modeling}

Non-AR prosody modeling refines prosody at the frame-level from time-dependent prosody features. In this process, the source-filter generator converts acoustic feature embeddings into the source representation, reflecting the prosody characteristics of the prompt by the attention mechanism and the modulation.

\subsection{Learning Objectives}
The learning objectives consist of three components: reconstruction loss, adversarial loss, and acoustic feature loss. The reconstruction loss is the L1 loss between the generated and the ground-truth’s mel-spectrogram. The adversarial loss utilizes LSGAN \cite{mao2017lsgan}, incorporating a multi-window discriminator \cite{chen2020hifisinger, ijcai2022synta} with 2D patch unit lengths $\{32,64,128\}$. The acoustic feature loss computes the sum of cross-entropy losses for each acoustic feature, comparing the output units of the prosody predictor with the ground-truth acoustic units.

\begin{figure}[t]
  \centering
  \centerline{\includegraphics[width=1.0\linewidth]{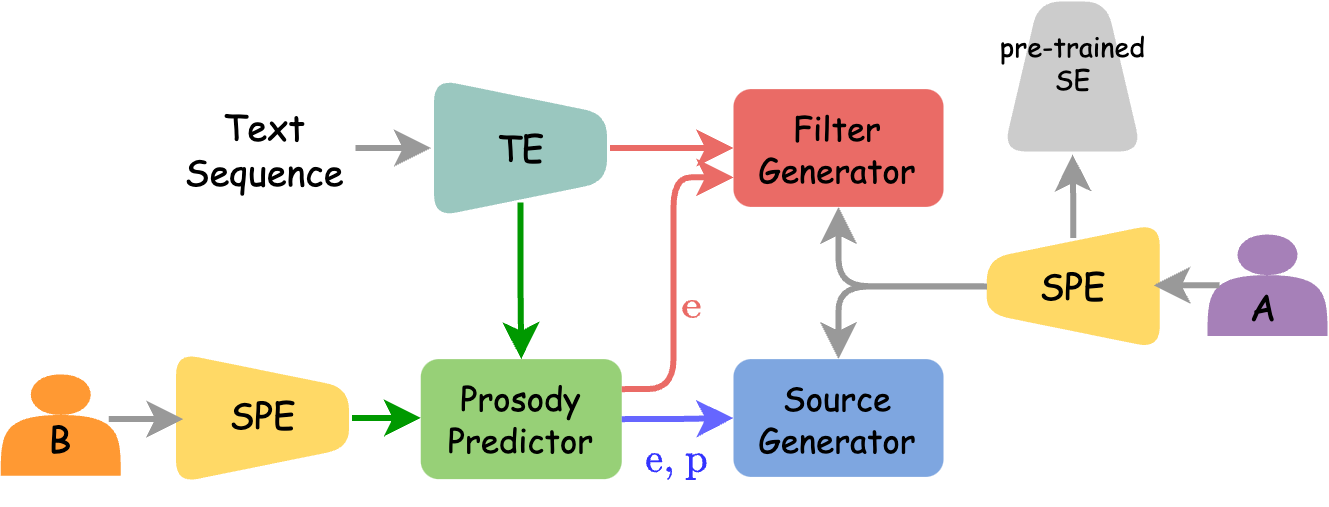}}
  \caption{Style transfer process to transfer speech style from speaker B to speaker A. The acoustic decoder is omitted for simplicity.}
  \label{fig:sst}
\end{figure}

\subsection{Multi-Task TTS}
The proposed model can perform multiple tasks according to the input condition. First, zero-shot TTS takes an unseen speaker’s prompt as input. Second, cross-lingual TTS is accomplished by using different languages for the speech prompt and input text. Additionally, the speech style transfer enters two different prompts into different modules, as illustrated in \cref{fig:sst}. These three tasks can be combined with each other. For example, zero-shot cross-lingual TTS or zero-shot style transfer, and even all three tasks can be performed at once, namely zero-shot cross-lingual speech style transfer. Detailed inference process are provided in \cref{app:details-inference}.

\section{Experimental Environments}
\subsection{Datasets}
\textbf{Training datasets}\space\space Training datasets consist of English and Korean speech datasets. We used the open datasets LibriTTS (train-clean-100, train-clean-360) \cite{libri} and VCTK \cite{vctk} and an internal dataset as the English dataset. As the Korean dataset, we used the open dataset AI-Hub\footnote{\url{https://aihub.or.kr}} with multi-style and an internal dataset. Specification for datasets are provided in \cref{app:data-information}. We re-sampled all speech data to a 22.05 kHz sampling rate, and obtained an 80-band mel-spectrogram as the acoustic feature. The Short-Time Fourier Transform (STFT) parameters included a bin size of 1024, with window size of 1024 and a hop sizes 256.

\textbf{Evaluation datasets}\space\space  We diversified the composition of the evaluation dataset. The dataset for evaluating zero-shot performance is made up of utterances from speakers not included in the training. Specifically, it is divided by language (English and Korean) and speaking style (neutral and expressive). For the English dataset, the neutral style is represented by LibriTTS dev-clean and the expressive style is represented by EXPRESSO \cite{nguyen2023expresso}. Four styles (confused, enunciated, happy, and sad) were selected from the various styles available in EXPRESSO. The Korean dataset includes the neutral style from the AI-Hub multi-speaker dataset and the expressive style from an internal dataset. The expressive internal dataset includes voices in emotional and theatrical styles.

\subsection{Experimental Setup}
\textbf{Baselines}\space\space  To evaluate speech generation performance under the same training data conditions, we trained GANSpeech \cite{yang21e_interspeech} and YourTTS \cite{casanova2022yourtts} as baselines. GANSpeech is a Non-autoregressive transformer-based multi-speaker TTS model, which adopts adversarial training using speaker conditioned discriminator for securing generalization. We configured GANSpeech to facilitate zero-shot and cross-lingual tasks by using a pre-trained speaker encoder \cite{chung2020in}, referred to as GANSpeech+. YourTTS, a conditional variational autoencoder-based end-to-end model, incorporates a learning objective for consistent speaker modeling and handles both cross-lingual and zero-shot tasks. We utilized the official implementation of YourTTS\footnote{\url{https://github.com/Edresson/YourTTS}}, but did not use the official pre-trained model of YourTTS because it has not been trained with Korean speech data. For evaluation of speech style transfer performance, we conducted a comparison with Daft-exprt \cite{zaidi22b_interspeech}, a speech style transfer model. Daft-exprt transfers style with extracted prosody information from the reference voice. We used the official implementation of Daft-exprt\footnote{\url{https://github.com/ubisoft/ubisoft-laforge-daft-exprt}}.

\textbf{Model configuration}\space\space The MultiVerse's encoders, generators, and decoder are all constructed with the transformer blocks. Except for the AR prosody predictor, all modules operate in a non-AR manner. 
The global style embedding is obtained from the output of an open-source pre-trained ResNet-based speaker recognition model \cite{chung2020in}. Detailed configurations for each module of MultiVerse and the corresponding hyper-parameters are described in \cref{app:details}.
Since the proposed model and the GANSpeech+ output mel-spectrogram, we utilized a pre-trained Avocodo \cite{bak2023avocodo} which is an universal neural vocoder. 
In the case of YourTTS, a neural vocoder was not employed because it directly generates waveforms.

\textbf{Training}\space\space  To obtain the phoneme sequence, we performed grapheme-to-phoneme (G2P) processing on the audio transcripts. For English, we utilized the IPA-based open-source tool\footnote{\url{https://github.com/rhasspy/gruut-ipa}}, and for Korean, an internal G2P tool was employed. The acoustic features included duration, fundamental frequency ($F_0$), and energy. Duration extraction employed the Montreal Forced Aligner 2.0 \cite{mcauliffe17_interspeech} and the internal Forced Aligner for English and Korean speech, respectively. Praat toolkit\footnote{\url{https://github.com/YannickJadoul/Parselmouth}} was used for $F_0$ extraction. Detailed acoustic feature pre-processing is described in the \cref{app:acoustic}. The training batch size was 48 and 8 NVIDIA A100 GPUs were used in training.
The optimization employed the ADAMW optimizer \cite{decoupled_weight_decay}, with the parameters $(\beta_1, \beta_2)$ set as $(0.8, 0.99)$, and a NOAM learning rate scheduler \cite{vaswani2017attention}. The training of the AR prosody predictor started at the peak learning rate to stabilize training. A random segment of the reference speech was selected as the prompt for each training iteration, and the reconstruction loss was not computed for that segment \cite{shen2023naturalspeech}. To prevent the model from excessively mimicking the prosody of the prompt, a shorter segment was used for the prosody predictor. MultiVerse and GANSpeech+ were trained for 600k iterations, while YourTTS was trained for 500k iterations.

\begin{table*}[t]
    \caption{The objective and subjective evaluation results in a zero-shot scenario using speech prompts with variations in style (Neutral (N) and Expressive (E)) and language (English (ENG) and Korean (KOR)).}
    \vskip -0.1in
    \centering
    \begin{small}
    \begin{tabular}{cclccccccc}
    \toprule
         \textbf{\thead{Prompt \\ Style}} & \textbf{\thead{Prompt \\ Language}} & \multicolumn{1}{c}{\textbf{Method}} &
        \textbf{\thead{CER\\($\downarrow$)}} & \textbf{\thead{WER\\($\downarrow$)}} & \textbf{\thead{SECS\\($\uparrow$)}} & \textbf{\thead{$F_0$ PCC\\($\uparrow$)}} & \textbf{N-MOS} & \textbf{S-MOS} & \textbf{PS-MOS} \\
        \midrule
         \multicolumn{10}{c}{\textbf{Intra-lingual}} \\ \midrule
         \multirow{8}{*}{N} & \multirow{4}{*}{ENG} & Ground-truth & \multicolumn{1}{r}{0.89} & \multicolumn{1}{r}{2.84} & \multicolumn{1}{r}{0.820} & \multicolumn{1}{r}{-} & \multicolumn{1}{r}{4.17$\pm$0.13} & \multicolumn{1}{r}{4.12$\pm$0.16} & \multicolumn{1}{r}{-} \\ \cline{3-10}
         ~ & ~ & GANSpeech+ & \multicolumn{1}{r}{\textbf{0.76}} & \multicolumn{1}{r}{\textbf{2.61}} & \multicolumn{1}{r}{0.736} & \multicolumn{1}{r}{0.037} & \multicolumn{1}{r}{3.83$\pm$0.15} & \multicolumn{1}{r}{3.72$\pm$0.18} & \multicolumn{1}{r}{-} \\
         ~ & ~ & YourTTS & \multicolumn{1}{r}{3.50} & \multicolumn{1}{r}{7.78} & \multicolumn{1}{r}{0.810} & \multicolumn{1}{r}{0.021} & \multicolumn{1}{r}{2.88$\pm$0.15} & \multicolumn{1}{r}{3.83$\pm$0.16} & \multicolumn{1}{r}{-} \\
         ~ & ~ & MultiVerse & \multicolumn{1}{r}{0.89} & \multicolumn{1}{r}{2.70} & \multicolumn{1}{r}{\textbf{0.852}} & \multicolumn{1}{r}{\textbf{0.073}} & \multicolumn{1}{r}{\textbf{3.87$\pm$0.14}} & \multicolumn{1}{r}{\textbf{4.42$\pm$0.12}} & \multicolumn{1}{r}{-} \\ \cline{2-10}
         ~ & \multirow{4}{*}{KOR} & Ground-truth & \multicolumn{1}{r}{4.15} & \multicolumn{1}{r}{21.08} & \multicolumn{1}{r}{0.845} & \multicolumn{1}{r}{-} & \multicolumn{1}{r}{4.15$\pm$0.11} & \multicolumn{1}{r}{4.63$\pm$0.07} & \multicolumn{1}{r}{-} \\ \cline{3-10}
         ~ & ~ & GANSpeech+ &  \multicolumn{1}{r}{\textbf{3.23}} & \multicolumn{1}{r}{\textbf{17.10}} & \multicolumn{1}{r}{0.740} & \multicolumn{1}{r}{0.069} & \multicolumn{1}{r}{\textbf{4.29$\pm$0.09}} & \multicolumn{1}{r}{3.19$\pm$0.12} & \multicolumn{1}{r}{-} \\
         ~ & ~ & YourTTS & \multicolumn{1}{r}{6.56} & \multicolumn{1}{r}{26.16} & \multicolumn{1}{r}{0.790} & \multicolumn{1}{r}{0.037} & \multicolumn{1}{r}{3.69$\pm$0.10} & \multicolumn{1}{r}{3.72$\pm$0.12} & \multicolumn{1}{r}{-} \\
         ~ & ~ & MultiVerse & \multicolumn{1}{r}{3.76} & \multicolumn{1}{r}{18.94} & \multicolumn{1}{r}{\textbf{0.834}} & \multicolumn{1}{r}{\textbf{0.147}} & \multicolumn{1}{r}{3.91$\pm$0.10} & \multicolumn{1}{r}{\textbf{4.12$\pm$0.11}} & \multicolumn{1}{r}{-} \\ \cline{1-10}
         \multirow{8}{*}{E} & \multirow{4}{*}{ENG} &Ground-truth & \multicolumn{1}{r}{1.52} & \multicolumn{1}{r}{6.37} & \multicolumn{1}{r}{0.806} & \multicolumn{1}{r}{-} & \multicolumn{1}{r}{3.71$\pm$0.15} & \multicolumn{1}{r}{4.45$\pm$0.11} & \multicolumn{1}{r}{4.26$\pm$0.15} \\ \cline{3-10}
         ~ & ~ & GANSpeech+ &  \multicolumn{1}{r}{\textbf{1.51}} & \multicolumn{1}{r}{\textbf{5.10}} & \multicolumn{1}{r}{0.700} & \multicolumn{1}{r}{0.064} & \multicolumn{1}{r}{\textbf{3.62$\pm$0.14}} & \multicolumn{1}{r}{3.30$\pm$0.20} & \multicolumn{1}{r}{3.17$\pm$0.21} \\
         ~ & ~ & YourTTS &  \multicolumn{1}{r}{4.49} & \multicolumn{1}{r}{10.62} & \multicolumn{1}{r}{0.755} & \multicolumn{1}{r}{0.047} & \multicolumn{1}{r}{2.99$\pm$0.16} & \multicolumn{1}{r}{3.76$\pm$0.16} & \multicolumn{1}{r}{3.51$\pm$0.18} \\ 
         ~ & ~ & MultiVerse & \multicolumn{1}{r}{1.79} & \multicolumn{1}{r}{6.05} & \multicolumn{1}{r}{\textbf{0.811}} & \multicolumn{1}{r}{\textbf{0.100}} & \multicolumn{1}{r}{3.27$\pm$0.16} & \multicolumn{1}{r}{\textbf{4.27$\pm$0.13}} & \multicolumn{1}{r}{\textbf{4.08$\pm$0.13}} \\  \cline{2-10}
         ~ & \multirow{4}{*}{KOR} &Ground-truth & \multicolumn{1}{r}{6.66} & \multicolumn{1}{r}{19.89} & \multicolumn{1}{r}{0.836} & \multicolumn{1}{r}{-} & \multicolumn{1}{r}{4.65$\pm$0.08} & \multicolumn{1}{r}{4.47$\pm$0.10} & \multicolumn{1}{r}{4.05$\pm$0.11} \\ \cline{3-10}
         ~ & ~ & GANSpeech+ & \multicolumn{1}{r}{\textbf{5.16}} & \multicolumn{1}{r}{\textbf{17.95}} & \multicolumn{1}{r}{0.733} & \multicolumn{1}{r}{0.028} &\multicolumn{1}{r}{\textbf{4.43$\pm$0.08}} & \multicolumn{1}{r}{3.48$\pm$0.13} & \multicolumn{1}{r}{3.78$\pm$0.11} \\
         ~ & ~ & YourTTS &  \multicolumn{1}{r}{8.07} & \multicolumn{1}{r}{24.77} & \multicolumn{1}{r}{0.793} & \multicolumn{1}{r}{0.027} & \multicolumn{1}{r}{3.96$\pm$0.10} & \multicolumn{1}{r}{4.06$\pm$0.10} & \multicolumn{1}{r}{4.05$\pm$0.10} \\
         ~ & ~ & MultiVerse & \multicolumn{1}{r}{5.22} & \multicolumn{1}{r}{18.54} & \multicolumn{1}{r}{\textbf{0.830}} & \multicolumn{1}{r}{\textbf{0.066}} & \multicolumn{1}{r}{4.28$\pm$0.08} & \multicolumn{1}{r}{\textbf{4.24$\pm$0.09}} & \multicolumn{1}{r}{\textbf{4.29$\pm$0.09}} \\  \midrule
         \multicolumn{10}{c}{\textbf{Cross-lingual}} \\ \midrule
          \multirow{6}{*}{N} & \multirow{3}{*}{\thead{ENG \\ (to KOR)}} & GANSpeech+ & \multicolumn{1}{r}{4.72} & \multicolumn{1}{r}{22.55} & \multicolumn{1}{r}{0.667} & \multicolumn{1}{r}{0.025} & \multicolumn{1}{r}{\textbf{4.20$\pm$0.08}} & \multicolumn{1}{r}{2.69$\pm$0.10} & \multicolumn{1}{r}{-} \\  
         ~ & ~ & YourTTS &  \multicolumn{1}{r}{10.40} & \multicolumn{1}{r}{36.07} & \multicolumn{1}{r}{0.780} & \multicolumn{1}{r}{0.011} & \multicolumn{1}{r}{3.04$\pm$0.10} & \multicolumn{1}{r}{3.32$\pm$0.11} & \multicolumn{1}{r}- \\ 
         ~ & ~ & MultiVerse &\multicolumn{1}{r}{\textbf{4.43}} & \multicolumn{1}{r}{\textbf{21.98}} & \multicolumn{1}{r}{\textbf{0.780}} & \multicolumn{1}{r}{\textbf{0.043}} & \multicolumn{1}{r}{4.09$\pm$0.09} & \multicolumn{1}{r}{\textbf{3.56$\pm$0.10}} & \multicolumn{1}{r}{-} \\ \cline{2-10} 
         ~ & \multirow{3}{*}{\thead{KOR \\ (to ENG)}} & GANSpeech+ &  \multicolumn{1}{r}{\textbf{0.28}} & \multicolumn{1}{r}{\textbf{1.18}} & \multicolumn{1}{r}{0.656} & \multicolumn{1}{r}{0.001} & \multicolumn{1}{r}{\textbf{3.72$\pm$0.15}} & \multicolumn{1}{r}{2.45$\pm$0.19} & \multicolumn{1}{r}- \\
         ~ & ~ & YourTTS & \multicolumn{1}{r}{5.46} & \multicolumn{1}{r}{10.14} & \multicolumn{1}{r}{0.761} & \multicolumn{1}{r}{0.015} & \multicolumn{1}{r}{2.60$\pm$0.15} & \multicolumn{1}{r}{\textbf{3.70$\pm$0.19}} & \multicolumn{1}{r}{-} \\
         ~ & ~ & MultiVerse & \multicolumn{1}{r}{0.39} & \multicolumn{1}{r}{1.34} & \multicolumn{1}{r}{\textbf{0.782}} & \multicolumn{1}{r}{\textbf{0.048}} & \multicolumn{1}{r}{3.63$\pm$0.14} & \multicolumn{1}{r}{3.12$\pm$0.19} & \multicolumn{1}{r}{-} \\  \cline{1-10} 
         \multirow{6}{*}{E} & \multirow{3}{*}{\thead{ENG \\ (to KOR)}} &GANSpeech+ & \multicolumn{1}{r}{4.99} & \multicolumn{1}{r}{23.28} & \multicolumn{1}{r}{0.637} & \multicolumn{1}{r}{0.089} & \multicolumn{1}{r}{\textbf{3.92$\pm$0.10}} & \multicolumn{1}{r}{2.68$\pm$0.11} & \multicolumn{1}{r}{3.15$\pm$0.12}  \\
         ~ & ~ & YourTTS & \multicolumn{1}{r}{10.22} & \multicolumn{1}{r}{34.71} & \multicolumn{1}{r}{0.735} & \multicolumn{1}{r}{0.072} & \multicolumn{1}{r}{2.90$\pm$0.11} & \multicolumn{1}{r}{3.08$\pm$0.12} & \multicolumn{1}{r}{3.16$\pm$0.11} \\
         ~ & ~ & MultiVerse &  \multicolumn{1}{r}{\textbf{4.44}} & \multicolumn{1}{r}{\textbf{22.45}} & \multicolumn{1}{r}{\textbf{0.758}} & \multicolumn{1}{r}{\textbf{0.122}} & \multicolumn{1}{r}{3.32$\pm$0.10} & \multicolumn{1}{r}{\textbf{3.20$\pm$0.11}} & \multicolumn{1}{r}{\textbf{3.56$\pm$0.11}} \\  \cline{2-10} 
         ~ & \multirow{3}{*}{\thead{KOR \\ (to ENG)}} & GANSpeech+ &  \multicolumn{1}{r}{0.47} & \multicolumn{1}{r}{\textbf{1.37}} & \multicolumn{1}{r}{0.645} & \multicolumn{1}{r}{0.060} & \multicolumn{1}{r}{3.79$\pm$0.15} & \multicolumn{1}{r}{2.61$\pm$0.19} & \multicolumn{1}{r}{3.21$\pm$0.18} \\ 
         ~ & ~ & YourTTS &  \multicolumn{1}{r}{6.31} & \multicolumn{1}{r}{11.46} & \multicolumn{1}{r}{0.763} & \multicolumn{1}{r}{0.053} & \multicolumn{1}{r}{2.47$\pm$0.15} & \multicolumn{1}{r}{\textbf{3.78$\pm$0.18}} & \multicolumn{1}{r}{\textbf{3.77$\pm$0.19}} \\ 
         ~ & ~ & MultiVerse &  \multicolumn{1}{r}{\textbf{0.42}} & \multicolumn{1}{r}{1.40} & \multicolumn{1}{r}{\textbf{0.773}} & \multicolumn{1}{r}{\textbf{0.083}} & \multicolumn{1}{r}{\textbf{3.84$\pm$0.13}} & \multicolumn{1}{r}{2.89$\pm$0.19} & \multicolumn{1}{r}{3.31$\pm$0.19} \\ 
    \bottomrule
    \end{tabular}
    \end{small}
    \vskip -0.1in
  \label{tab:sub-results}
\end{table*}

\textbf{Metrics}\space\space For objective evaluation, we measured speech intelligibility, speaker and prosody similarity between the prompt and the synthesized speech. The evaluation of speech intelligibility involves comparing the Character- and Word-error-rate (CER, WER) measured by automatic speech recognition (ASR) using the pre-trained Whisper\footnote{\url{https://huggingface.co/openai/whisper-large}} model \cite{whisper}. The speaker embedding cosine similarity (SECS) using open-source voice encoder\footnote{\url{https://github.com/resemble-ai/Resemblyzer}} measures speaker similarity between the prompt and the synthesized speech. For prosody similarity evaluation, $F_0$ Pearson correlation coefficient ($F_0$ PCC) between $F_0$ of the prompt and synthesized speech is calculated.
For evaluating style transfer performance, we measured the similarity in pitch and duration between the synthesized speech and the style prompt. Dynamic Time Warping was employed to measure the $F_0$ similarity ($F_0$ DTW). SECS was also used to evaluate similarity between the target speaker’s audio sample and the style-transferred speech. Subjective evaluation included three Mean Opinion Score (MOS) tests: The Naturalness-MOS (N-MOS) test evaluates the naturalness and intelligibility of the speech. The Similarity-MOS (S-MOS) test assesses the speaker similarity and the Prosody Similarity-MOS (PS-MOS) test evaluates the prosody similarity between the synthesized and the prompt speech for expressive style speech only. A total of 10 native speakers evaluated 15 English audio samples, and 24 native speakers evaluated 10 Korean audio samples. A detailed description about the MOS tests are described in \cref{app:ablation}.

\section{Experimental Results}
\subsection{Zero-shot Scenario} \label{zero-shot scenario}
We conducted objective and subjective evaluations for speech synthesis using unseen speaker's or unseen language's speech prompt. We categorized experiments into intra-lingual and cross-lingual experiments based on the language of the speech prompt and the target language. 
For the evaluation, 400 audio samples were synthesized using speech prompt and unpaired input text. Both the speech prompt and the text were selected randomly from the evaluation datasets. The speech prompt consisted of randomly sliced audio segments, each lasting 3 seconds.
Experimental results are presented in \cref{tab:sub-results}, respectively\footnote{Audio samples are available at \url{https://nc-ai.github.io/speech/publications/multiverse/index.html}.}. In \cref{app:ablation}, we additionally conducted an ablation study to assess the individual impact of proposed methods used in MultiVerse.

\textbf{Intra-lingual}\space\space The evaluation results show that the MultiVerse outperforms across languages and speech styles. In particular, it synthesized speech maintaining speaker or prosody similarity of the prompt. Lower SECS and higher $F_0$ PCC results indicate that the robust style and prosody modeling of MultiVerse enables it to generate speech highly similar to the prompt, even in expressive styles. Subjective evaluation results also confirmed that the proposed model generates speech that is aurally natural and similar to the prompt. Notably, it excels in speaker and prosody similarity compared to baselines. In the naturalness test (N-MOS), GANSpeech+ achieves slightly higher scores than the proposed model. However, due to the low generalization performance of GANSpeech+, it synthesized speech with the speaking style and speaker identity far from prompt speech. The similarity tests (S- and PS-MOS) indicate that GANSpeech+ relies on learned features rather than reflecting information from the prompt speech. Meanwhile, YourTTS achieves higher speaker similarity than GANSpeech+, but suffers from lower quality. These results show that both baseline models have limitations in similarity and speech robustness.

\textbf{Cross-lingual}\space\space The evaluation results of MultiVerse’s cross-lingual task show a similar tendency to the intra-lingual task. The S-MOS and PS-MOS scores in `KOR (to ENG)', which synthesizes English speech by inputting a Korean expressive prompt, are relatively lower than `ENG (to KOR)' because of a data imbalance caused by the absence of expressive style in the English training dataset. However, when amount of expressive style speech in the English training datasets are prepared, similarity performance may improve even under the combination of unseen language, speaker, and style conditions. In N-MOS, MultiVerse occasionally received lower scores than GANSpeech+. We speculate that this may be due to the disparity between the language of the prompt and the input text. GANSpeech+ generally exhibits high naturalness but synthesizes speech with low similarity because of synthesizing speech with low speaker similarity and a neutral style, ignoring the prompt.
\begin{table}[t]
    \centering
    \caption{The results of subjective evaluation for the comparison with data-driven large-scale models.}
    \vskip -0.1in
    \begin{small}
    \begin{tabular}{cccc}
    \toprule
        \textbf{\thead{Prompt \\ Language}} & \textbf{Model} & \textbf{N-MOS} &\textbf{S-MOS} \\
        \midrule
        \multirow{3}{*}{\thead{ENG}} & \multicolumn{1}{l}{Ground-truth} & \multicolumn{1}{r}{4.31$\pm$0.11} & \multicolumn{1}{r}{4.21$\pm$0.14} \\ \cline{2-4}
        ~ & \multicolumn{1}{l}{VALL-E} & \multicolumn{1}{r}{3.56$\pm$0.14} & \multicolumn{1}{r}{\textbf{4.36$\pm$0.11}}  \\
        ~ & \multicolumn{1}{l}{MultiVerse} & \multicolumn{1}{r}{\textbf{3.85$\pm$0.14}} & \multicolumn{1}{r}{4.30$\pm$0.13} \\ \midrule
        \multirow{2}{*}{\thead{CHN \\ (to ENG)}} & \multicolumn{1}{l}{VALL-EX} & \multicolumn{1}{r}{3.05$\pm$0.27} & \multicolumn{1}{r}{3.09$\pm$0.31} \\
        ~ & \multicolumn{1}{l}{MultiVerse} & \multicolumn{1}{r}{\textbf{3.51$\pm$0.26}} & \multicolumn{1}{r}{\textbf{3.33$\pm$0.30}} \\
    \bottomrule
    \end{tabular}
    \end{small}
  \label{tab:large-results}
\end{table}
\begin{table}[t]
    \caption{The results of objective evaluation for the comparison with data-driven models. The training data size in hours for each model is indicated in parentheses.}
    \vskip -0.1in
    \begin{small}
    \begin{tabular}{lrrrr}
    \toprule
    \multicolumn{1}{c}{\textbf{Method}} & \multicolumn{1}{c}{\textbf{CER}} & \multicolumn{1}{c}{\textbf{WER}} & \multicolumn{1}{c}{\textbf{SECS}} & \multicolumn{1}{c}{\textbf{\thead{$F_0$ \\ PCC}}} \\ \midrule
    \multicolumn{5}{c}{\textbf{Intra-lingual}} \\ \midrule
Mega-TTS (20k)                   & 0.000                             & 0.000                             & 0.800                             & \textbf{0.269}                                  \\
MultiVerse & 0.000 & 0.000 & \textbf{0.835}  & 0.215                                  \\  \cline{1-5}

NaturalSpeech2 (44k)           & 0.002                            & 0.008                            & 0.814                             & 0.079                                  \\ 
MultiVerse                 & 0.002                            & 0.008                            & \textbf{0.826}                             & \textbf{0.091}                                  \\ \cline{1-5}
Voicebox (60k)                   & 0.016                            & 0.052                            & \textbf{0.779} & 0.076                                  \\
MultiVerse                 & \textbf{0.014}                            & \textbf{0.048}  & 0.718 & \textbf{0.187}                                  \\ \midrule
\multicolumn{5}{c}{\textbf{Cross-lingual}} \\   \midrule
Mega-TTS (20k)                   &\textbf{0.007}                            & \textbf{0.058}                            & \textbf{0.747}                             & 0.095                                  \\ 
MultiVerse                 & 0.013                            & 0.077                            & 0.681                             & \textbf{0.164}                                  \\ \cline{1-5}
Voicebox (50k) &\textbf{0.001}&\textbf{0.013}                            & \textbf{0.812}                            & 0.063     \\
MultiVerse & 0.002 & 0.017 & 0.692 & \textbf{0.130}  \\
    \bottomrule
    \end{tabular}
    \end{small}
    \vskip -0.1in
  \label{tab:large-results-obj}
\end{table}

\begin{table*}[t]
    \centering
    \caption{Evaluation results of speech style transfer.}
    \begin{small}
    \begin{tabular}{cccccccc}
    \toprule
        \textbf{Task} & \textbf{Model} & \textbf{$F_0$ DTW($\downarrow$)} & \textbf{Dur. RMSE($\downarrow$)} & \textbf{CER($\downarrow$)} & \textbf{WER($\downarrow$)} & \textbf{SECS($\uparrow$)} & \textbf{N-MOS} \\
        \midrule
        \multirow{2}{*}{same-text} & \multicolumn{1}{l}{Daft-Exprt} & \multicolumn{1}{r}{0.370} & \multicolumn{1}{r}{3.480} & \multicolumn{1}{r}{8.03} & \multicolumn{1}{r}{24.71} & \multicolumn{1}{r}{0.859} & \multicolumn{1}{r}{2.63$\pm$0.21} \\
        ~ & MultiVerse & \multicolumn{1}{r}{\textbf{0.348}} & \multicolumn{1}{r}{\textbf{1.500}} & \multicolumn{1}{r}{\textbf{3.25}} & \multicolumn{1}{r}{\textbf{17.94}} & \multicolumn{1}{r}{\textbf{0.865}} & \multicolumn{1}{r}{\textbf{3.69$\pm$0.22}} \\ \midrule
        \multirow{2}{*}{different-text} & \multicolumn{1}{l}{Daft-Exprt} & \multicolumn{1}{r}{\textbf{0.438}} & \multicolumn{1}{r}{-} & \multicolumn{1}{r}{3.51} & \multicolumn{1}{r}{14.02} & \multicolumn{1}{r}{0.862} & \multicolumn{1}{r}{2.82$\pm$0.21} \\
        ~ & MultiVerse & \multicolumn{1}{r}{0.440} & \multicolumn{1}{r}{-} & \multicolumn{1}{r}{\textbf{2.27}} & \multicolumn{1}{r}{\textbf{11.55}} & \multicolumn{1}{r}{\textbf{0.868}} & \multicolumn{1}{r}{\textbf{3.27$\pm$0.18}} \\
    \bottomrule
    \end{tabular}
    \end{small}
  \label{tab:sst-results}
\end{table*}

\subsection{Comparison with Data-Driven Models}
We compared performance between our proposed model and data-driven large-scale models. We selected VALL-E \cite{wang2023valle} and VALL-EX \cite{zhang2023vallex} as baselines for subjective evaluation. VALL-E and VALL-EX were trained on over 60k hours of English speech data and over 70k hours of English and Chinese speech data, respectively. For a fair comparison, we synthesized speech using MultiVerse based on identical prompts and scripts used in publicly available speech samples of each model\footnote{\url{https://www.microsoft.com/en-us/research/project/vall-e-x/}}. \cref{tab:large-results} presents the results of the N-MOS and S-MOS tests. Despite using relatively small amounts of training data, MultiVerse is able to synthesize speech that is not only more natural but also comparable in similarity to the prompt when compared to large-scale models. Moreover, even though MultiVerse was not exposed to a Chinese dataset during training, it outperforms VALL-EX. For objective evaluation, Mega-TTS \cite{jiang2023mega}\footnote{\url{https://mega-tts.github.io/demo-page/}}, NaturalSpeech2 \cite{shen2023naturalspeech}\footnote{\url{https://speechresearch.github.io/naturalspeech2/}}, and Voicebox \cite{voicebox}\footnote{\url{https://voicebox.metademolab.com/}} were selected as baselines. In the cross-lingual task, Mega-TTS used Chinese prompts, and Voicebox used Spanish, French, German, Portuguese, and Polish prompts to generate English speech in the style of the prompts. \cref{tab:large-results-obj} describes the results which demonstrate that MultiVerse achieved comparable zero-shot performance to large-scale TTS models with only about 1.2k hours of training data. Notably, the F0 PCC results confirm that MultiVerse can generate voices with higher prosody similarity. However, in the cross-lingual task, MultiVerse generated speech with lower speaker similarity than the baselines. We speculate that the performance degradation is due to the prompt speech being in a language not included in the training data.

\subsection{Speech Style Transfer}
\cref{tab:sst-results} presents objective and subjective evaluation results for scenarios in which the input text aligns with the content of the style prompt (same-text) and scenarios in which the target text differs from the one of the style prompt (different-text). All models were trained using Korean datasets, and the full frame of the prompt speech was employed during inference. Objective evaluations were conducted on a total of 225 audio samples, and the naturalness of 10 audio samples of each model was assessed by 9 Korean participants in the listening test. Object evaluation results demonstrate that MultiVerse achieves speech style transfer in both same- and different-text environments. Please note that MultiVerse can model prosody suitable to input text and prompt speech by in-context learning, while Daft-exprt takes extracted acoustic features for transferring speech style. Nevertheless, MultiVerse effectively transfers the style of prompt, especially rhythm and speed. The SECS result demonstrates that MultiVerse, which utilize speaker information from prompt speech, preserves speaker information similarly to Daft-exprt, which uses deterministic speaker ID. In the N-MOS test, participants rated MultiVerse as generating more natural and clear speech than Daft-exprt. It is attributed to low intelligibility and restricted style reflection on the input text, as Daft-exprt directly transfers style using extracted acoustic features.
\subsection{Acoustic Feature Modeling} \label{acoustic_modeling}

\begin{figure}[t]
  \centering
  \centerline{\includegraphics[width=1.0\linewidth]{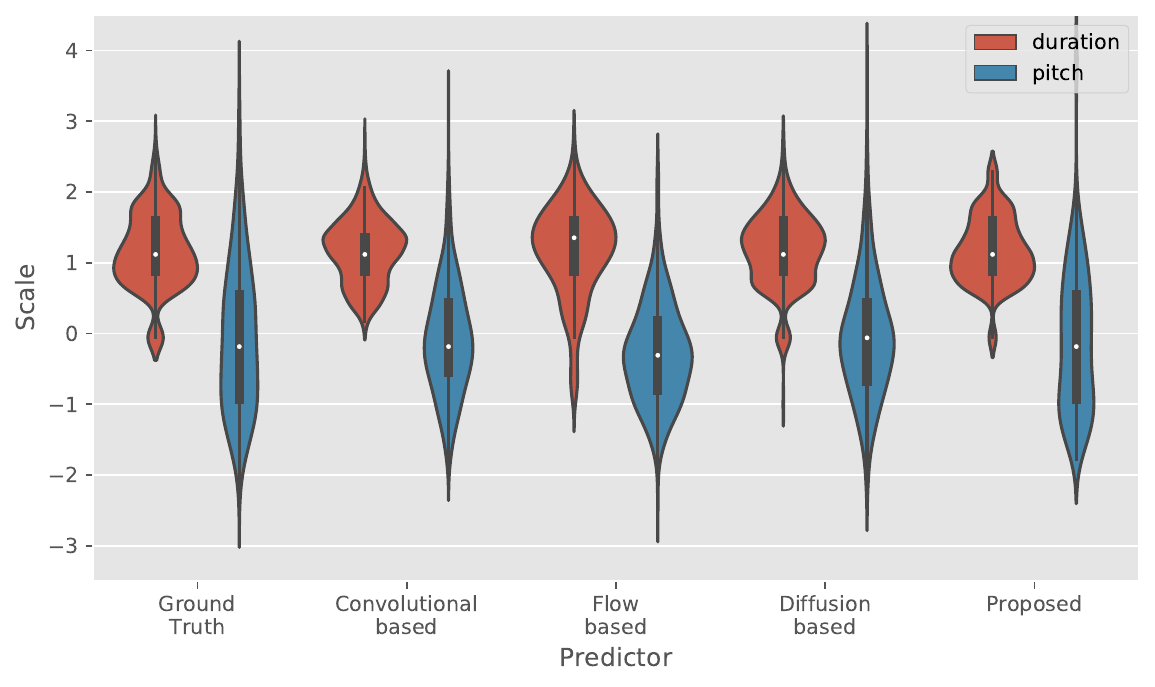}}
  \caption{Violin plot describing duration and pitch distributions.}
  \label{fig:pr-distribution2}
  \vskip -0.2in
\end{figure}

\begin{figure}[t]
  \centering
 \centerline{\includegraphics[width=1.1\linewidth]{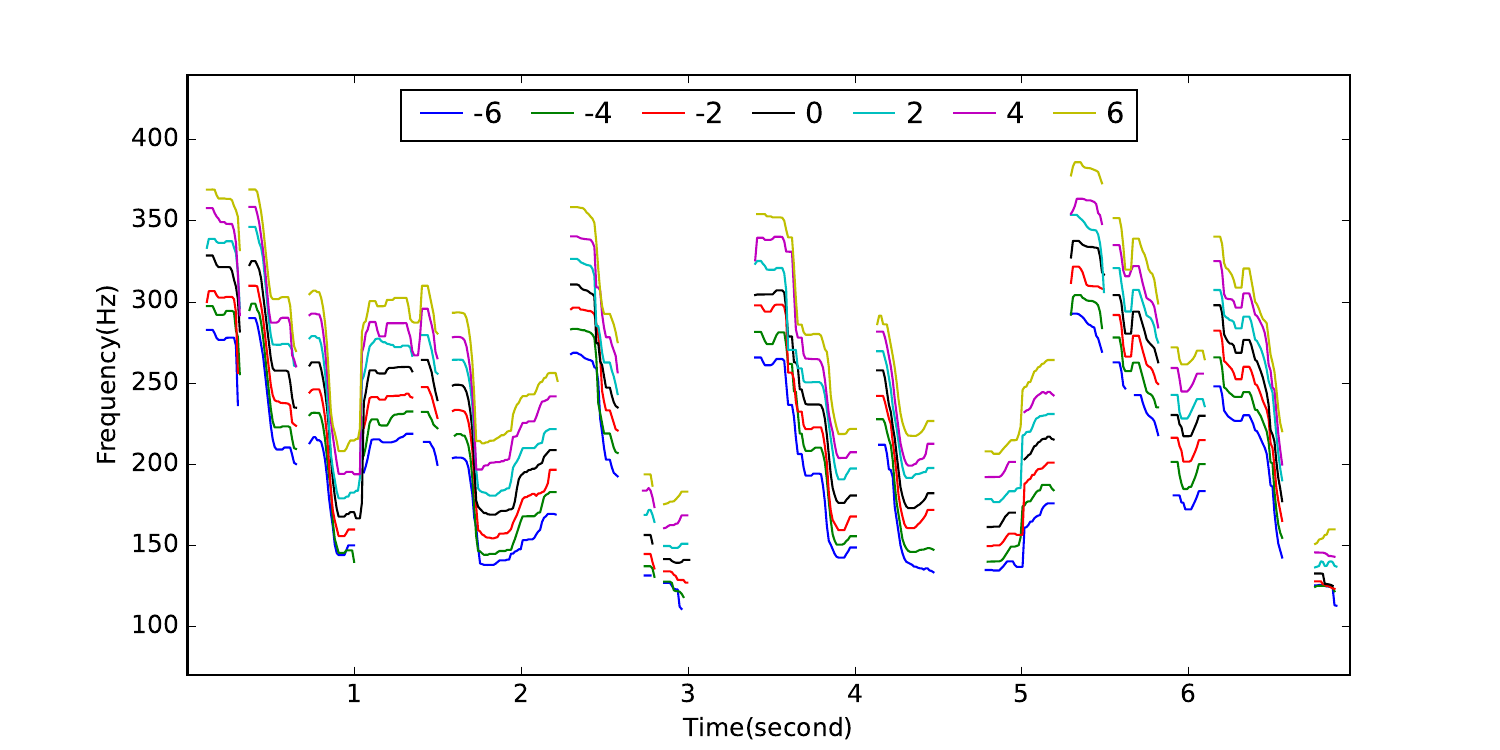}}
  \caption{$F_0$ contours of pitch-shifted synthesized speech whose predicted pitch units are manipulated with $\{-6, -4, -2, 2, 4, 6\}$.  }
  \vskip -0.2in
  \label{fig:pr-contour}
\end{figure}

We compared the acoustic feature modeling performance of the AR prosody predictor with convolutional layer-based, Flow-based and Denoising Diffusion Probabilistic Model-based predictors. Specifically, we selected the convolutional layer-based predictor from FastSpeech2 \cite{fastspeech2}, the Flow-based duration predictor and pitch predictor from YourTTS and VarianceFlow \cite{varianceflow}, and DDPM-based predictor from \citet{li23j_interspeech}, respectively. Each predictor was trained with Korean data. The distributions of predicted duration and pitch according to each predictor are illustrated in \cref{fig:pr-distribution2}. Acoustic features of expressive speech have a distribution close to multi-modal or has large variation. The convolutional layer and Flow-based predictors that do not consider the time-dependency, underfit close to the average. Diffusion based predictor also fail to model the target distribution, while the AR prosody predictor approximates the target distribution relatively closely. It indicates that the prompt-based AR structure effectively models the time-dependent characteristics of prosody. Additionally, the proposed AR prosody predictor can control prosody by adjusting acoustic feature indices. \cref{fig:pr-contour} illustrates the $F_0$ contour of synthesized speech generated by manipulating pitch index sequences to constant values. The change in the acoustic index corresponds to a variation in the $F_0$ contour of the synthesized speech.

\section{Conclusion}

This paper introduces MultiVerse, an efficient and expressive zero-shot multi-task TTS system designed to address the limitations of existing zero-shot TTS systems that depend on large-scale training datasets. MultiVerse employs a structure that disentangles speech components into filter and source representations: this structure contributes to achieving zero-shot TTS performance comparable to data-driven TTS approaches, even with a small amount of data. Additionally, it enhances prosody similarity through a hybrid prosody modeling that combines both autoregressive and non-autoregressive mechanisms. Quantitative and qualitative evaluations across various language and style demonstrate that MultiVerse excels in zero-shot, cross-lingual TTS, and speech style transfer. 

\section{Limitations}
MultiVerse, which generates mel-spectrograms, requires a system to convert the mel-spectrograms into waveforms, such as a neural vocoder. Therefore, the performance of the vocoding system can potentially affect the performance of MultiVerse. Additionally, utilizing pre-processed acoustic features, i.e., duration, pitch, and energy, becomes more costly as the amount of training data increases. Hence, one of the next goals of this research could be to incorporate unsupervised modeling methods for acoustic features.

\section{Broader Impacts}
We aimed to enhance the versatility of deep-learning-based TTS models. While speech generative models offer valuable support for creating digital content, concerns arise about their potential misuse for fraud and crime. This study is designed to minimize these potential negative impacts and effectively support TTS models for content creators. We emphasize ethical considerations, especially regarding data privacy, by ensuring that all voice data used in training and evaluation is sourced from publicly available datasets or internal datasets with the explicit consent of the speakers. Moreover, the study looks ahead to future societal implications, striving to expand the capabilities of TTS models in a manner that aligns with ethical and social responsibilities related to content creation.

\section{Acknowledgements}
This research was supported by Culture, Sports and Tourism R\&D Program through the Korea Creative Content Agency grant funded by the Ministry of Culture, Sports and Tourism in 2024 (Project Name: Development of Co-Pilot technology for automatic completion of generative AI-based 3D Webtoon, Project Number: RS-2024-00400004, Contribution Rate: 25\%)
\bibliography{main}

\appendix

\begin{figure*}[t]
    \vskip 0.2in
    \centering
    \centerline{\includegraphics[width=0.7\linewidth]{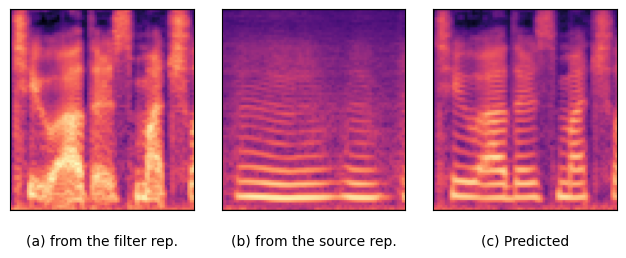}}
    \caption{Mel-spectrograms of synthesized audio sample: (a) 
 generated from the filter representation only. (b) generated from the source representation only. (c) generated from the coarse mel-representation. }
    \label{fig:app-mel_rep}
\end{figure*}

\begin{table*}[t!]
\centering
\caption{Analysis results on filter and source representations.}
\vskip 0.15in
\label{tab:analysis_content_style}
\begin{tabular}{@{}cccccc@{}}
\toprule
  \textbf{Task} &
  \textbf{Model} &
  \textbf{CER($\downarrow$)} &
  \textbf{WER($\downarrow$)} &
  \textbf{SECS($\uparrow$)} &
  \textbf{$F_0$ PCC($\uparrow$)} \\ \midrule
\multirow{3}{*}{\begin{tabular}[c]{@{}c@{}}Intra-\\ lingual\end{tabular}} &
  \multicolumn{1}{l}{MultiVerse} &
  \textbf{3.11} &
  \textbf{12.02} &
  \textbf{0.831} &
  \textbf{0.111} \\ \cline{2-6}
 &
  \multicolumn{1}{l}{filter representation} &
  3.26 &
  13.23 &
  0.668 &
  0.023 \\
 &
  \multicolumn{1}{l}{source representation} &
  - &
  - &
  0.574 &
  0.090 \\ \midrule
\multirow{3}{*}{\begin{tabular}[c]{@{}c@{}}Cross-\\ lingual\end{tabular}} &
  \multicolumn{1}{l}{MultiVerse} &
  \textbf{2.65} &
  \textbf{12.42} &
  \textbf{0.775} &
  \textbf{0.089} \\ \cline{2-6}
 &
  \multicolumn{1}{l}{filter representation} &
  3.03 &
  14.56 &
  0.622 &
  0.020\\
 &
  \multicolumn{1}{l}{source representation} &
  - &
  - &
  0.544 &
  0.074 \\ \bottomrule
\end{tabular}
\vskip -0.1in
\end{table*}

\section{Related Works}
\label{Related works}

\textbf{Zero-shot TTS}\space\space Zero-shot TTS synthesizes speech for unseen speakers 
not present in the training dataset. Learning general speech features, like speaker identity and speaking style, is crucial in this task. Zero-shot TTS models often incorporate a pre-trained speaker encoder for modeling speaker information \cite{jia2018transfer, kumar2021normalization, cooper2020zero}. Some employ specific objectives to improve speaker similarity \cite{casanova21b_interspeech, casanova2022yourtts}. However, challenges arise when generating voices significantly different from the training data, impacting similarity and naturalness. Recent advances in language models have prompted exploration of data-driven methods in speech synthesis \cite{wang2023valle, borsos2023audiolm, kharitonov2023speak, shen2023naturalspeech}, enhancing generalization to unseen voices. Despite this, acquiring large speech datasets is costly and challenges persist in obtaining diverse data for various languages. Existing models have primarily focused on speaker similarity, leaving the issue of prosody similarity unresolved.

\textbf{Cross-lingual TTS}\space\space Cross-lingual TTS aims to generate speech in a language that is different from the monolingual speaker while preserving speaker's voice. However, training on multilingual multi-speaker data may entangle language and speaker information, resulting in diminished similarity. Therefore, disentangling the language from the speaker becomes crucial in cross-lingual TTS. \citet{zhang19e_interspeech, xin2020cross} propose adversarial layers to disentangle speaker and language information. \citet{xin2021disentangled} utilize mutual information minimization to decouple the information. Despite the objective of disentanglement, these models suffer from unstable training, creating a trade-off between disentanglement and speaker similarity \cite{zhang19e_interspeech}. Recently, data-driven methods have also been applied to improve the generalization in cross-lingual TTS \cite{zhang2023vallex}. 

\textbf{Speech style transfer}\space\space Speech style transfer aims to synthesize speech with a speaking style similar to the reference speech, notwithstanding differences like identity or content. Style transfer models learn to model inherent elements of voice style, disentangling these elements from content and speaker identity \cite{lee21h_interspeech, zaidi22b_interspeech, huang2022generspeech, yuan2020improving}. \citet{lee21h_interspeech} disentangles acoustic features (duration, prosody, energy) by encoding them separately. Daft-Exprt \cite{zaidi22b_interspeech} uses domain adversarial training to separate prosody and speaker information. \citet{huang2022generspeech} proposes mix-style layer normalization to remove style information from filter representation. While these studies enhance style transfer performance, their limited disentanglement hinders style transfer in out-of-domain environments \cite{sigurgeirsson2023prosody}.

\begin{figure*}[t]
    \vskip 0.2in
  \centering
  \centerline{\includegraphics[width=0.5\linewidth]{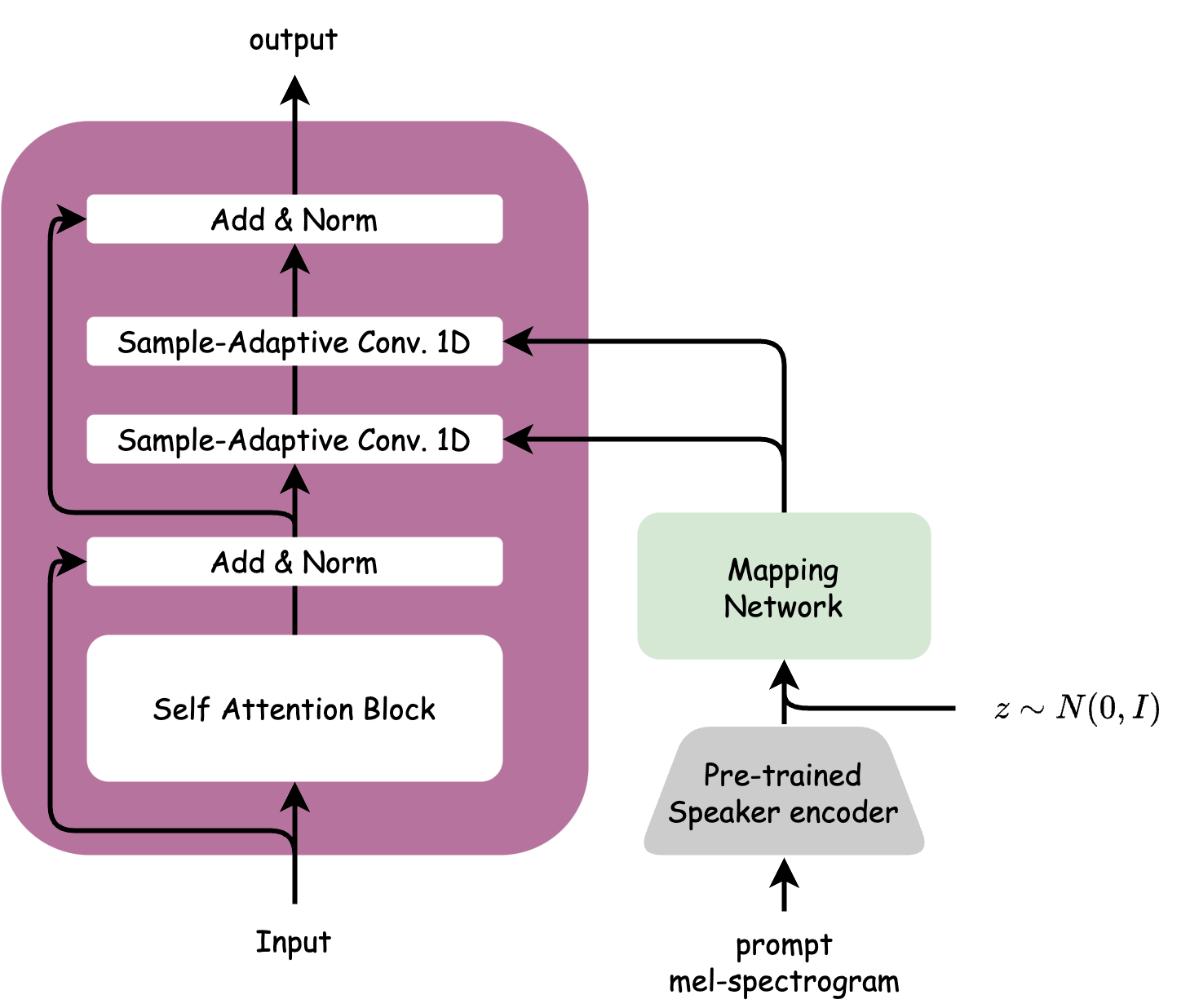}}
  \caption{The diagram of the proposed acoustic decoder with the sample-adaptive kernel selection.}
  \label{fig:sample-adaptive}
  \vskip -0.2in
\end{figure*}

\begin{figure*}[t]
    \vskip 0.2in
    \centering
    \subfigure[MultiVerse w/o sample-adaptive kernel selection.]{\includegraphics[width=0.45\linewidth]{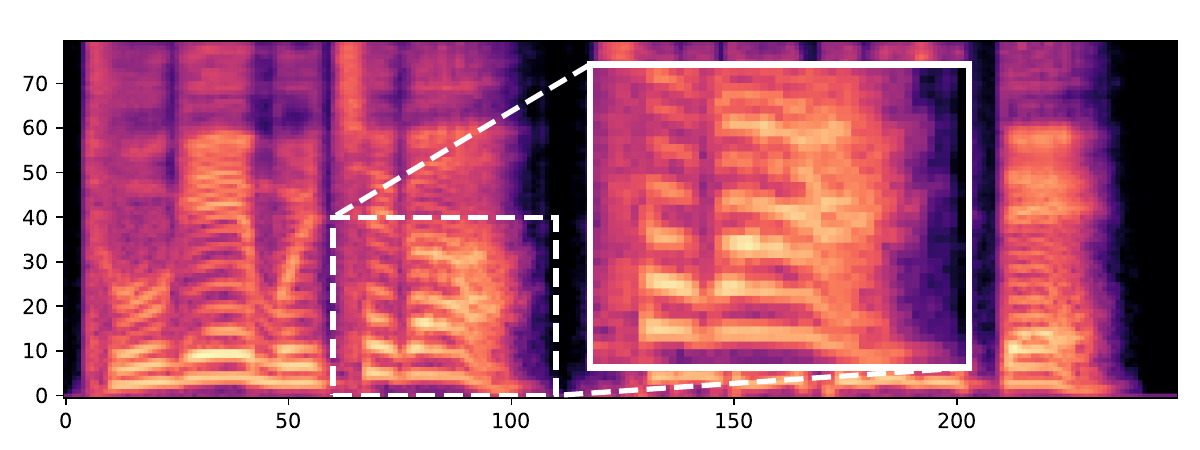}} 
    \subfigure[MultiVerse.]{\includegraphics[width=0.45\linewidth]{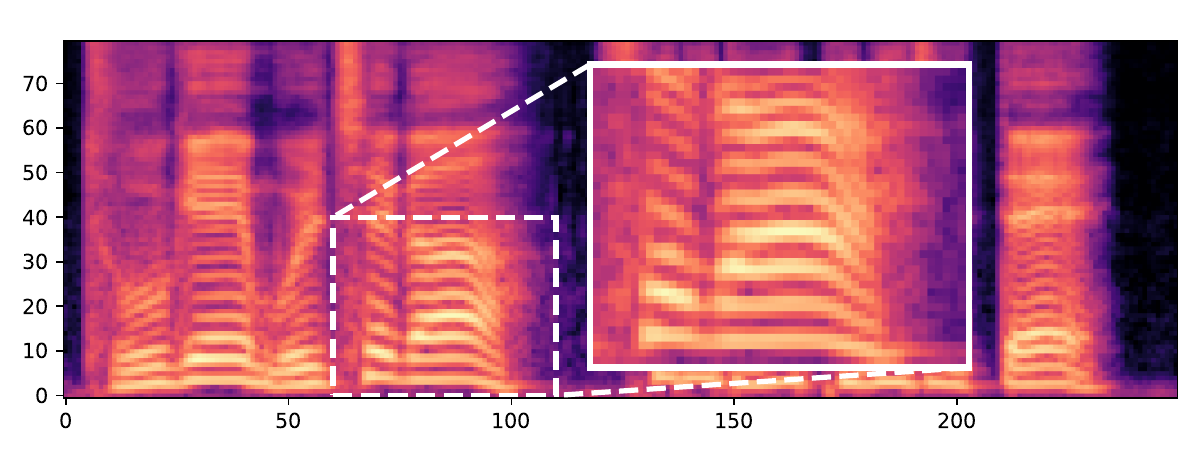}}
    \caption{Mel-spectrograms of synthesized audio sample: MultiVerse and MultiVerse w/o sample-adaptive kernel selection.}
    \label{fig:ablation-sample-adaptive}
    \vskip -0.2in
\end{figure*}

\section{Analysis on Filter and Source Representations}\label{app:representations}



MultiVerse adopts source-filter theory-based decomposed modeling to learn disentangled representations which are divided into filter and source. As described in \cref{proposed_methods}, the filter generator produces the filter representation related to speech content, pronunciation and accent. On the other hand, the source generator produces the source representation that contains prosodic information that is less correlated to the content, such as intonation, rhythm, and stress patterns.

\cref{fig:app-mel_rep} presents mel-spectrograms generated from the each representations. To analyze these representations in detail, we conducted objective evaluation. In this experiments, audio samples were generated by passing both representations individually through the acoustic decoder, not forming the coarse mel-representation. \cref{tab:analysis_content_style} demonstrates the evaluation results.
CER and WER results indicate that the synthesized speech from the filter representation has comparable intelligibility to that generated by MultiVerse. This means that the filter representation primarily contains linguistic information of input text. Additionally, SECS result shows that the filter representation is more related to the speaker identity than the source representation. 
Meanwhile, the synthesized speech from the source representation is sounding like mumble sounds ``Hmmmm, Mmmmmm ..."; ASR model failed to recognize the speech.
 However, it has more similar pitch distribution than that from the filter representation because the source representation, generated from acoustic features, learns the prosodical patterns included in the prompt speech.

\section{Detailed on Acoustic Decoder}\label{app:sample-adaptive-detail}
The proposed acoustic decoder employs sample-adaptive kernel selection \cite{gigagan} to learn convolutional filters suitable for the speech prompt. This approach generates a mel-spectrogram while preserving the information of the coarse mel-representation. \cref{fig:sample-adaptive} depicts the detailed structure of sample-adaptive kernel selection.

\begin{figure*}[t]
    \vskip 0.2in
  \centering
  \centerline{\includegraphics[width=1.0\linewidth]{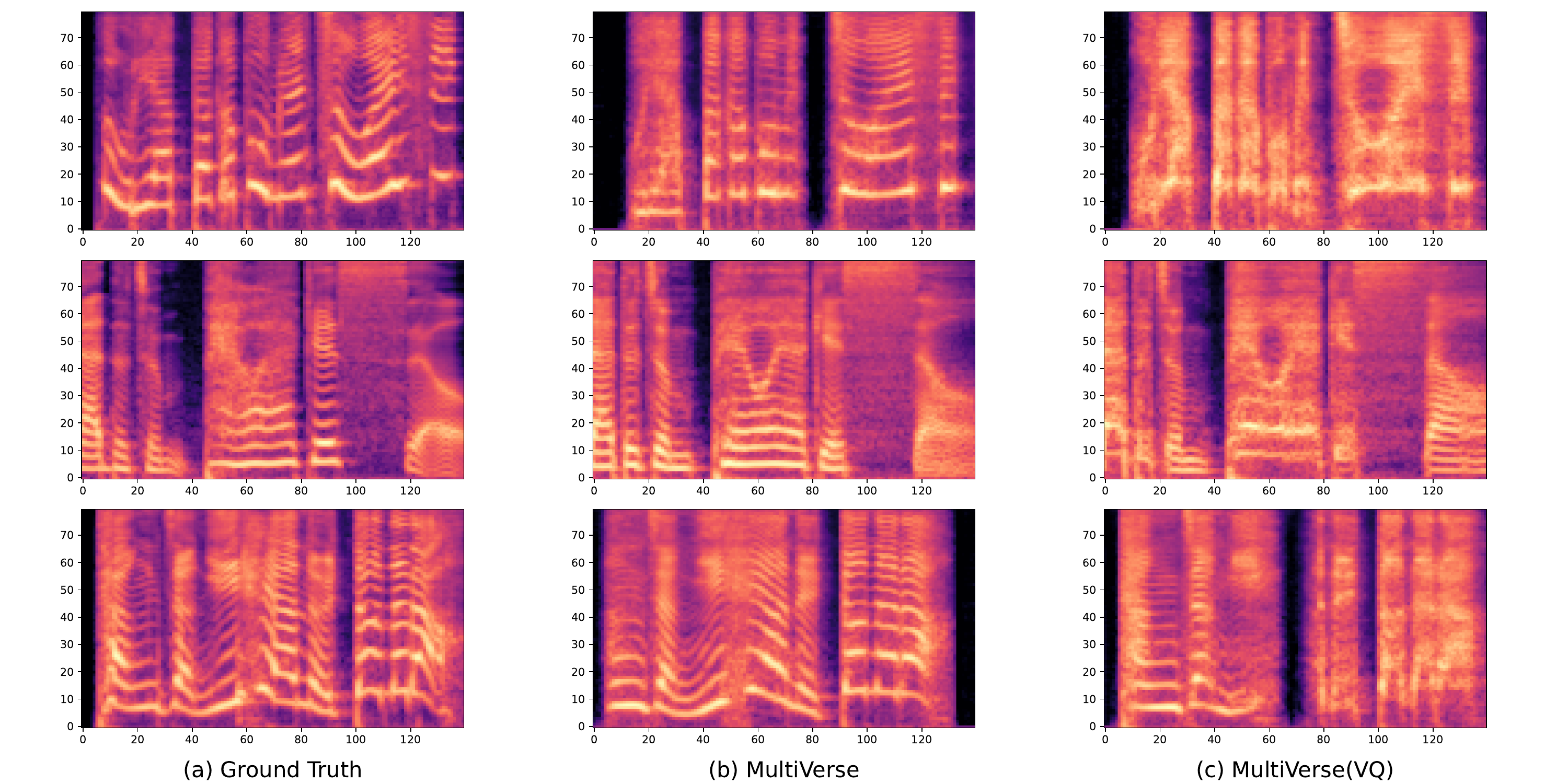}}
  \caption{Mel-spectrograms of (a) ground-truth, (b) generated speech from MultiVerse and (c) generated speech from MultiVerse with vector quantization (VQ) based prosody modeling.}
  \label{fig:pllm}
  \vskip -0.2in
\end{figure*}
The specific process is as follows: the mapping network maps the style vector from a global style embedding and random noise sampled from a normal Gaussian distribution. The K-bank convolutional filters of each sample-adaptive convolution layer are aggregated by a weighted sum based on the weights predicted for each kernel by the style vector. Subsequently, the aggregated filter undergoes modulation and demodulation by scale, where the scale is obtained from the output of an affine layer with the style vector as input \cite{stylegan2}.

The proposed acoustic decoder replaces the convolutional layer of the feed-forward transformer block with a sample-adaptive kernel selection-based convolutional layer. ReLU activation function is used for non-linearity between the two sample-adaptive convolutional layers. Detailed parameter descriptions for sample-adaptive kernel selection are provided in \cref{tab:Hyper-parameter}.

\section{Comparison with Vector Quantization based Prosody Modeling}\label{app:pllm}
The P-LLM of Mega-TTS also leverages autoregressive language model-based prosody modeling. However, as mentioned in \cref{ar_prosody}, the approach using vector quantization-based codebooks is influenced by the quantity of training data. To verify this, a simple comparison was conducted. We trained a model, referred as MultiVerse(VQ), by replacing MultiVerse's prosody modeling with Mega-TTS's VQ encoder. Both MultiVerse and MultiVerse(VQ) synthesized speech from the test dataset, using ground-truth mel-spectrogram as prompts and ground-truth phoneme durations. This experimental setup aimed to observe the performance of vector quantization-based models with limited training data.

We observed that the synthesized speech by MultiVerse(VQ) occasionally resulted in distorted audio, as depicted in \cref{fig:pllm}. MultiVerse(VQ) produced more distorted speech when targeting expressive style. It also fails to reflect the prosody of the prompt speech, indicating that expressive style was not effectively modeled by vector quantization, likely due to the relatively small number of expressive style audio samples. In contrast, MultiVerse demonstrated relatively robust synthesis of expressive style speech.

\begin{table}[t]
\caption{Detailed dataset information.}
\centering
\label{tab:table-data-information}
\begin{tabular}{lcrr}
\toprule
\textbf{Dataset} & \textbf{Language} & \multicolumn{1}{l}{\textbf{\thead{Number of \\ Speakers}}} & \multicolumn{1}{l}{\textbf{Time(h)}} \\
\hline
\multicolumn{1}{l}{LibriTTS}         & ENG           & 1133                                            & \multirow{3}{*}{262}                 \\
\multicolumn{1}{l}{VCTK}            & ENG           & 101                                             &                                      \\
Internal         & ENG           & 42                                              &                                      \\
\hline
AiHub            & KOR            & 46                                              & \multirow{2}{*}{969}                 \\
Internal         & KOR            & 229                                             &  \\                                    
\bottomrule
\end{tabular}
\end{table}


\section{Detailed Dataset Information}
\label{app:data-information}

\cref{tab:table-data-information} describes detailed dataset information. The English training set was constructed from approximately 262 hours of speech data. The LibriTTS \cite{libri}, VCTK \cite{vctk}, and internal datasets were recorded from 1133, 101, and 42 speakers, respectively\footnote{License: CC-BY-4.0}. The speech styles included both neutral narration and conversational styles. The Korean dataset consisted of a total of 969 hours of speech data, with the AI-Hub multi-speaker\footnote{\url{https://www.aihub.or.kr/aihubdata/data/view.do?currMenu=115&topMenu=100&aihubDataSe=data&dataSetSn=542}} and internal datasets recorded from 46 and 229 speakers, respectively. It included various speech styles, such as narration and acting. 10\% of all datasets were excluded from the training set for testing purposes.

\section{Inference Details}\label{app:details-inference}
In this section, we provide the inference process in detail. For zero-shot TTS, both input text and a prompt voice are required. In this case, a mismatch between the text and prompt is permissible. Whether the input text and prompt voice's language match determines if it is intra-lingual or cross-lingual TTS. At first, the input text is processed to obtain the output $x$ from the text encoder. Next, the prompt speech is used to obtain the hidden representation $r$ from the speech prompt encoder, and the global style embedding $r_g$ is obtained from the pre-trained speaker encoder. The concatenated $x$ and $r$ along the time axis serve as a prefix for the AR prosody predictor to autoregressively decode the phoneme acoustic features, including duration $\mathbf{d}$, pitch $\mathbf{p}$, and energy $\mathbf{e}$. The predicted phoneme-specific duration is utilized for upsampling to the frame level. The filter generator outputs the filter representation using the upsampled version of $x$ and $\mathbf{e}$. Simultaneously, the source generator produces the source representation from the upsampled version of $\mathbf{p}$ and $\mathbf{e}$, with $r$ utilized in generating both representations. These two representations are combined and transformed into a mel-spectrogram by the acoustic decoder, where $r_g$ is used to determine the decoder's filter.

For style transfer, input text and two audio samples are required: prompt speech for the target speaker, denoted as $y_s$, and prompt speech for the target style, represented as $y_p$. The flow of generation is the same as in zero-shot TTS, but the application of prompts differs. From the speech prompt encoder, we obtain the hidden representation for the speaker $r_s$ and the hidden representation for style $r_p$ for each prompt speech. The global style embedding is obtained from $y_s$. $r_p$ is utilized for predicting phoneme acoustic features, while $r_s$ is used for applying the prompt. Consequently, it is possible to generate a voice that reflects the style of $y_p$ with the speaker identity of $y_s$. Moreover, prosody-controllable TTS is facilitated by controlling the index of the predicted acoustic features.

\begin{table*}[ht!]
\centering
\caption{Detailed hyper-parameters of MultiVerse.}
\label{tab:Hyper-parameter}
\begin{tabular}{@{}ccc@{}}
\toprule
                    \textbf{Module}     &  \textbf{Configuration}   & \textbf{Value}               \\ \midrule
\multirow{5}{*}{Text Encoder}               & Encoder Layers        & 6\\
                                            & Feed-forward dim      & 2048                \\
                                            & Hidden dim            & 512                 \\
                                            & Kernel size           & 3                   \\
                                            & Number of heads              & 8                   \\ \midrule
\multirow{5}{*}{Speech Prompt Encoder}      & Encoder Layers        & 3                   \\
                                            & Feed-forward dim      & 2048                \\
                                            & Hidden dim            & 512                 \\
                                            & Kernel size           & 9                   \\
                                            & Number of heads              & 8                   \\ \midrule
\multirow{7}{*}{AR Prosody Predictor}       & Encoder Layers        & 3                   \\
                                            & Feed-forward dim      & 2048                \\
                                            & Hidden dim            & 512                 \\
                                            & Number of heads       & 8              \\
                                            & Duration codebook, dim    & 32, 512             \\
                                            & Pitch codebook, dim  & 64, 512             \\
                                            & Energy codebook, dim & 64, 512             \\ \midrule
\multirow{5}{*}{Filter / Source Generator}  & Encoder Layers        & 3                   \\
                                            & Feed-forward dim      & 2048                \\
                                            & Hidden dim            & 512                 \\
                                            & Kernel size           & 3                   \\
                                            & Number of heads       & 8                   \\ \midrule
\multirow{5}{*}{Acoustic Decoder}           & Encoder Layers        & 3                   \\
                                            & Feed-forward dim      & 2048                \\
                                            & Hidden dim            & 512                 \\
                                            & Kernel size           & 3                   \\
                                            & Number of heads       & 8                   \\ \midrule
\multirow{5}{*}{Sample-Adaptive Kernel Selection} & Mapping network depth & 4
                                            \\
                                            & Mapped style dim & 256
                                            \\
                                            & Noise dim & 64
                                            \\
                                            & Global style dim & 512
                                            \\
                                            & Size of kernel bank  & 4
                                            \\ \midrule
\multirow{4}{*}{Multi-Window Discriminator} & Number of discriminators     & 3                   \\
                                            & Window size           & 32, 64, 128         \\
                                            & Conv2d size           & 3                   \\
                                            & Hidden size           & 128                 \\ \midrule
\multicolumn{2}{c}{Total Number of Parameters}                      
         & 260.62M \\ \bottomrule
\end{tabular}
\vskip -0.2in
\end{table*}

\section{Model Configuration}\label{app:details}
Table \ref{tab:Hyper-parameter} describes detailed information of hyper-parameters of modules in MultiVerse. 

\section{Acoustic Feature Pre-Processing} \label{app:acoustic}
To obtain sequences of acoustic units for training, acoustic features, such as duration, fundamental frequency ($F_0$), and energy, were pre-processed using the following procedures. Each acoustic feature sequence was calculated, followed by normalization and quantization, to be transformed into acoustic unit sequences. Specifically, the procedures for each acoustic feature is as follows: For duration, to obtain the duration per phoneme, we used both the internal forced aligner and an external aligner \cite{mcauliffe17_interspeech} for Korean and English, respectively. The duration sequences were already in integer values, so we used them directly as duration unit sequences without additional normalization and quantization. We set the maximum duration value to 32. $F_0$ extraction from the speech signal utilized a Praat-based extractor\footnote{\url{https://github.com/YannickJadoul/Parselmouth}}. The $F_0$ sequences, extracted in Hertz units, were averaged per phoneme, then were normalized using speaker-specific statistical information. The normalized $F_0$ sequence values were quantized into 64 values within a certain range to obtain the pitch unit sequence. In this paper, the normalized $F_0$ sequence was clipped to the range from -4 to +4. For energy, frame-wise energy of the speech signal was calculated as the magnitude of the linear spectrogram and averaged per phoneme. The normalized energy sequence was quantized into 64 values within the range of -5 to +5.

\section{Details on the Subjective Evaluation}\label{app:subject}
\begin{figure*}
    \centering
    \subfigure[Screen setting for evaluation of N-MOS.]{\includegraphics[width=0.95\linewidth]{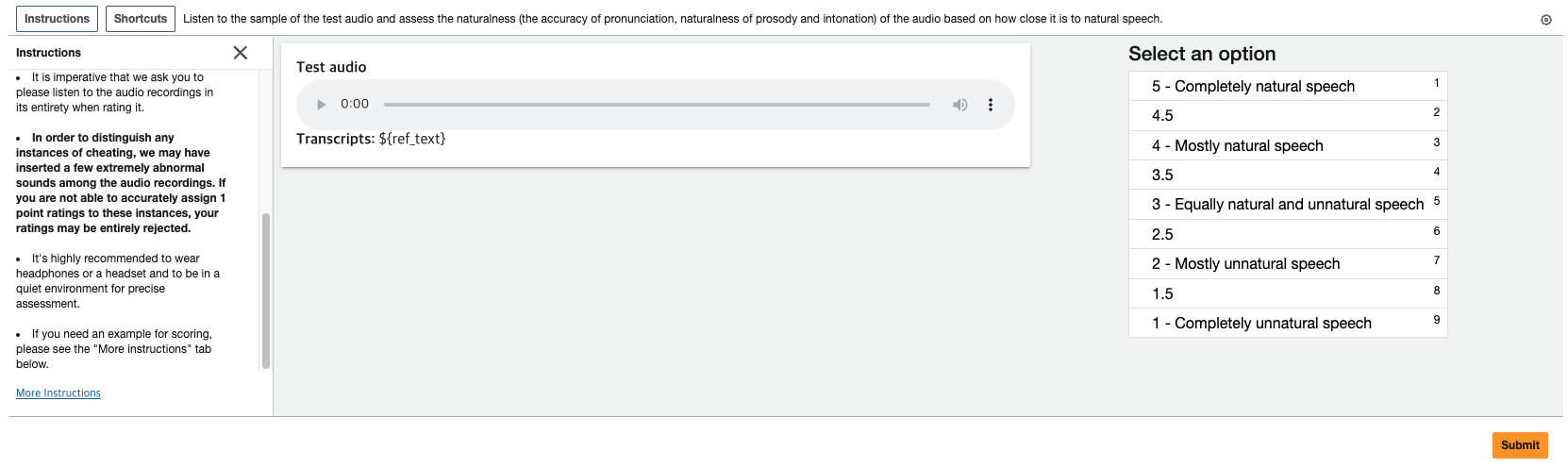}} 
    \subfigure[Screen setting for evaluation of S-MOS.]{\includegraphics[width=0.95\linewidth]{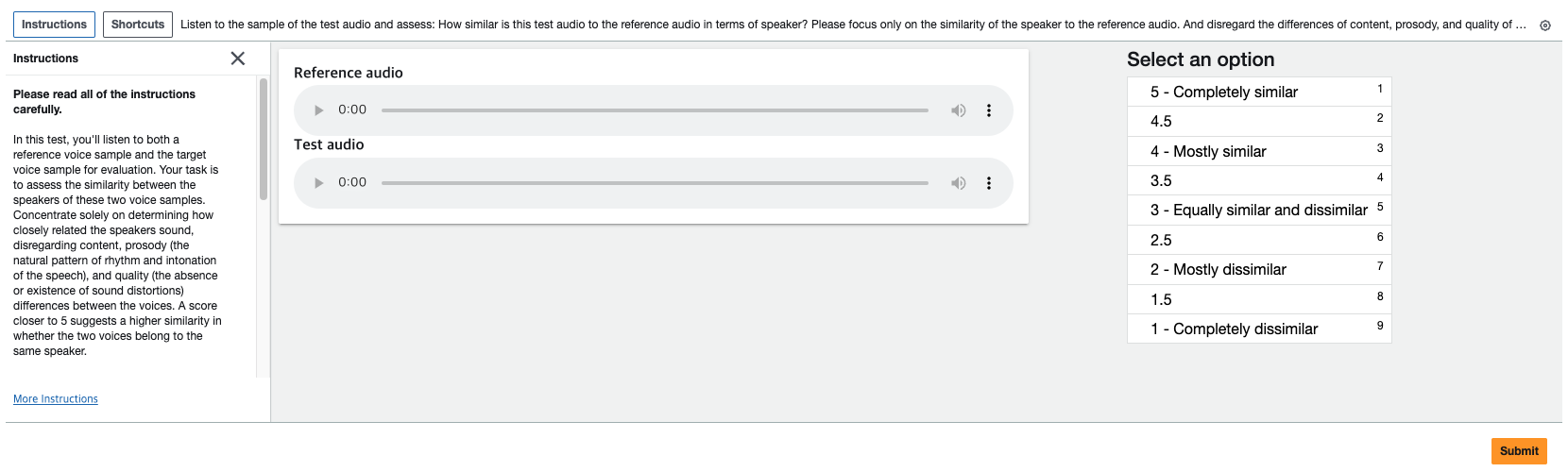}} 
    \subfigure[Screen setting for evaluation of PS-MOS.]{\includegraphics[width=0.95\linewidth]{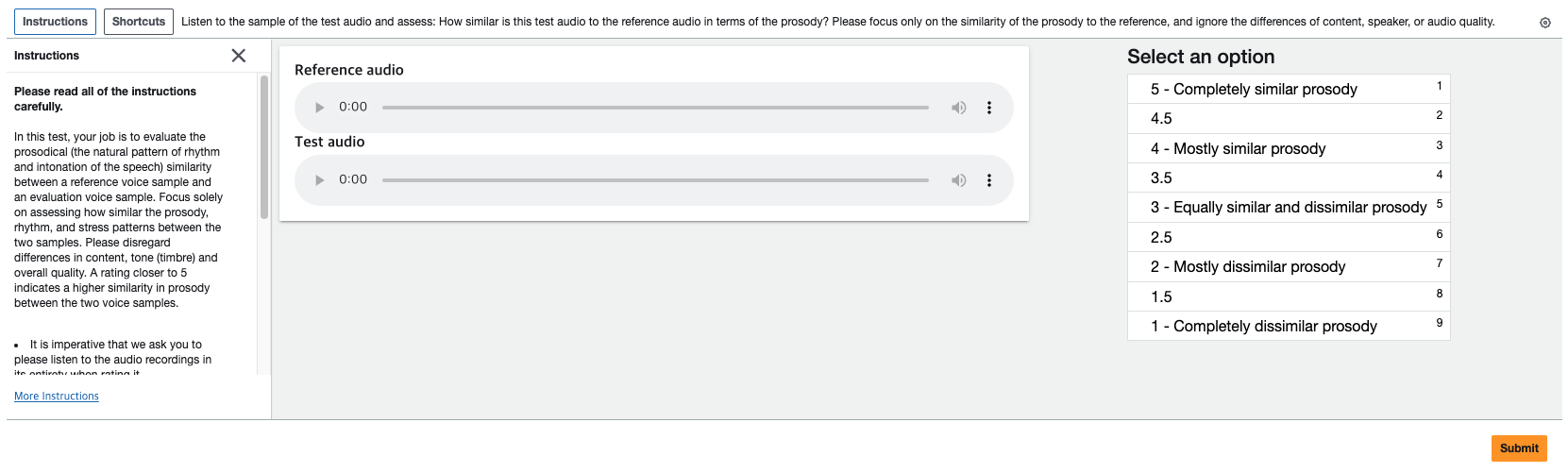}}
    \caption{Screen settings for evaluation of N-MOS, S-MOS, and PS-MOS.}
    \label{fig:MOS_screen}
\end{figure*}

For the subjective evaluation of samples, we performed three kinds of listening tests: the naturalness (N-MOS), speaker similarity (S-MOS), and prosody similarity (PS-MOS). Amazon Mechanical Turk (MTurk)\footnote{\url{https://www.mturk.com/}} and internal evaluation tools were used for the evaluation of the English and Korean samples, respectively. For each task (intra-lingual neutral/expressive, cross-lingual neutral/expressive), 15 test samples for English and 10 test samples for Korean were randomly selected per model. 10 native English participants and 24 native Korean participants rated the audio samples. 
Evaluation scores were evaluated at 0.5-point intervals from 1 to 5 points. 
We guided participants to evaluate audio samples by focusing only on the evaluation factor while ignoring other factors. In N-MOS test, the quality of the sound is ignored and only the naturalness of the speech is evaluated. For S-MOS test, we emphasized that participants should concentrate only on determining how closely related synthesized speech and prompt speech, disregarding content and prosody. Meanwhile, we requested participants to assess how similar the prosody, including rhythm and stress patterns, between the synthesized and prompt speech, disregarding content and timbre.
The actual evaluation screen and contents used are shown in Figure \ref{fig:MOS_screen}.

\begin{table*}[htbp!]
\centering
\caption{The result of the ablation study.}
\vskip 0.15in
\label{tab:ablation}
\begin{tabular}{@{}ccccc@{}}
\toprule
  \textbf{Task} &
  \textbf{Model} &
  \textbf{CER($\downarrow$)} &
  \textbf{WER($\downarrow$)} &
  \textbf{SECS($\uparrow$)} \\ \midrule
\multirow{5}{*}{\begin{tabular}[c]{@{}c@{}}Intra-\\ Lingual\end{tabular}} &
  \multicolumn{1}{l}{MultiVerse} &
  \textbf{3.11} &
  \textbf{12.02} &
  \textbf{0.831} \\ \cline{2-5}
 &
  \multicolumn{1}{l}{w/o source-filter} &
  2.78 &
  11.07 &
  0.818 \\
 &
  \begin{tabular}[l]{@{}c@{}}w/o sample-adaptive kernel selection\end{tabular} &
  
  3.01 &
  11.53 &
  0.817 \\
 &
  \multicolumn{1}{l}{w/o FiLM layer} &
  3.26 &
  11.83 &
  0.817 \\  \midrule
\multirow{5}{*}{\begin{tabular}[c]{@{}c@{}}Cross-\\ Lingual\end{tabular}} &
  \multicolumn{1}{l}{MultiVerse} &
  \textbf{2.65} &
  \textbf{12.42} &
  \textbf{0.775} \\ \cline{2-5}
 &
  \multicolumn{1}{l}{w/o source-filter} &
  2.49 &
  12.15 &
  0.760 \\
 &
  \begin{tabular}[l]{@{}c@{}}{w/o sample-adaptive kernel selection}\end{tabular} &
  2.66 &
  12.47 &
  0.760 \\
 &
  \multicolumn{1}{l}{w/o FiLM layer} &
  2.65 &
  12.34 &
  0.758 \\ \bottomrule
\end{tabular}
\end{table*}

\section{Ablation Study} \label{app:ablation}
To examine the specific impacts of the proposed methods, we compared models by sequentially removing individual components from the baseline architecture of our proposed model, namely the source-filter, sample-adaptive kernel selection, and the FiLM layer. Since the AR prosody predictor has already been compared with other predictors in \cref{acoustic_modeling}, we did not include it here. Evaluations were conducted for both intra- and cross-lingual tasks, similar to the zero-shot scenario. CER, WER, and SECS were used as evaluation metrics. The ablation evaluation results are detailed in \cref{tab:ablation}.

The evaluation results for CER and WER demonstrate that the FiLM layer, among the proposed methods, enhances speech robustness. No other method, excluding the FiLM layer, improved speech robustness. On the other hand, performance improved when the source-filter method was not used, which is related to the trade-off between speaker similarity and speech robustness, as observed in the objective evaluation results in \cref{zero-shot scenario}. All proposed methods influenced the enhancement of speaker similarity. Even when only one of the proposed methods was not utilized, SECS deteriorated. This indicates that the proposed methods contribute to improving the MultiVerse's generalization performance to learn speaker characteristics. We also observed that sample-adaptive kernel selection helps the acoustic decoder to generate a higher-quality mel-spectrogram. The difference in mel-spectrograms between MultiVerse with and without sample-adaptive kernel selection is depicted in \cref{fig:ablation-sample-adaptive}. MultiVerse without sample-adaptive kernel selection synthesized mel-spectrogram with decreased quality, resulting in lower intelligibility or the presence of artifacts.

\end{document}